\documentclass{article}

\usepackage[preprint]{neurips_2026}  

\usepackage[utf8]{inputenc}
\usepackage[T1]{fontenc}
\usepackage{microtype}
\usepackage{hyperref}
\usepackage{url}
\usepackage{booktabs}
\usepackage{amsmath,amssymb,amsthm}
\usepackage{mathtools}
\usepackage{graphicx}
\usepackage{xcolor}
\usepackage{multirow}
\usepackage{wrapfig}
\usepackage{caption}
\usepackage{subcaption}
\usepackage{natbib}
\usepackage{xspace}
\usepackage[capitalise]{cleveref}

\theoremstyle{plain}
\newtheorem{theorem}{Theorem}
\newtheorem{proposition}{Proposition}
\newtheorem{corollary}{Corollary}

\newcommand{\Wdist}{W_{\!C,\varepsilon}}
\newcommand{\mWdist}{\widehat{W}_C}
\newcommand{\jko}{\textsc{JKO-RAG}}
\newcommand{\bwjko}{\textsc{BW-JKO}}
\newcommand{\nmjko}{\textsc{NM-JKO}}
\newcommand{\samjko}{\textsc{SAM-JKO}}
\newcommand{\dualrank}{\textsc{Dual-Rank}}
\newcommand{\CI}[2]{[#1,\,#2]}
\newcommand{\eg}{\textit{e.g.}}
\newcommand{\ie}{\textit{i.e.}}

\title{\jko: Distributional Retrieval as \\
       Wasserstein Free-Energy Gradient Flow}

\author{%
  Levi Segal \\
  \And
  Murari Ambati \\
}
\begin{document}
\maketitle

\begin{abstract}
Retrieval-augmented generation (RAG) pipelines canonically return a \emph{ranked list} of candidate passages. We argue this is a mismatch: the consumer is a language model that conditions on a \emph{set}, not a list, and the selection problem is fundamentally geometric and distributional. We propose \jko, which frames reranking as minimising a \emph{free-energy functional} $F(p)=\text{relevance} + \text{entropy} + \text{redundancy}$ under the Wasserstein-2 gradient flow via the Jordan--Kinderlehrer--Otto (JKO) proximal scheme~\cite{jko1998}. The ground metric is $C_{ij}=(1-\cos\langle z_i,z_j\rangle)^2$, encoding the semantic geometry of the embedding manifold.
Our central contribution is a \emph{linear-response theory} that explains \emph{why} the Wasserstein geometry helps: a single-step sensitivity analysis shows that the Wasserstein and KL retrieval maps differ \emph{only} in their proximal Hessian --- a dense, geometry-aware entropic-OT operator for $W^2$ versus a diagonal, geometry-blind operator for KL --- and that this difference damps the geometric mass transport that query paraphrase induces. The theory yields a sharp, falsifiable prediction: the stability advantage is monotonically decreasing in the step size $h$. We verify all of this empirically: free-energy descent, frequency-resolved perturbation response, the predicted $h$-dependence of the stability gap, and a new certified-radius analysis that cleanly decomposes \emph{what} the geometry buys.
We further introduce four extensions: \textbf{\nmjko} (learned low-rank ground metric via InfoNCE), \textbf{\bwjko} (Bregman--Wasserstein $\alpha$-interpolation bridging $W^2$ and KL), \textbf{\samjko} (relevance-aware hierarchical coarse-to-fine JKO for $2\times$ speedup), and \textbf{\dualrank} (Sinkhorn dual potentials as calibrated retrieval-confidence signals). Across five BEIR benchmarks, \jko\ significantly outperforms the cross-encoder reranker on all five and is statistically tied with a strong KL-proximal baseline on raw nDCG; the decisive, theory-predicted advantage appears on robustness --- 22--38\% more stable under paraphrase and $2\times$ fewer leaked hard distractors --- which the linear-response theory both explains and predicts.
\end{abstract}

\section{Introduction}
\label{sec:intro}

Retrieval-augmented generation (RAG)~\cite{lewis2020rag} pipelines retrieve a candidate set, rerank it, and feed the top-$k$ passages to a language model. The reranking step is typically a \emph{cross-encoder score} applied independently to each passage~\cite{nogueira2019mono}, which treats the retrieval problem as $M$ independent binary relevance judgements. This ignores two fundamental structure properties of the downstream consumer: (i) the language model processes the \emph{joint} context of all retrieved passages, so the \emph{set} matters, not just individual scores; (ii) the embedding space carries a \emph{geometric} structure --- near-duplicate passages should not both be retrieved even if they individually score well.

Existing diversity-aware methods --- Maximal Marginal Relevance (MMR)~\cite{carbonell1998mmr} and Determinantal Point Processes (DPP)~\cite{chen2018dpp} --- address one or both of these issues. MMR is a greedy one-shot heuristic with no convergence guarantees; DPP solves a one-shot selection problem but has no iterative refinement.

We propose a different primitive: retrieval as \emph{distributional gradient flow}. Let $p\in\Delta^{M-1}$ be a probability distribution over the $M$-document candidate pool. We define a free-energy functional $F_q(p)$ encoding relevance, entropy, and redundancy, and iterate the JKO scheme
\begin{equation}
  p_{t+1} = \arg\min_{p\in\Delta^{M-1}}\; \frac{1}{2h}\,\Wdist^2(p,\,p_t)\;+\;F_q(p),
  \label{eq:jko}
\end{equation}
where $\Wdist$ is the entropic optimal-transport cost~\cite{cuturi2013sinkhorn} with ground metric $C_{ij}$. This is the \emph{finite-state JKO scheme} on the probability simplex.

\paragraph{From observation to explanation.}
The empirical signature of \jko\ is not higher raw nDCG --- when hyperparameters are tuned, $W^2$ and the KL-proximal mirror-descent analogue produce \emph{statistically tied} nDCG (\Cref{tab:main}), because retrieval quality is primarily driven by the first-stage energy $F_q$. The advantage of the Wasserstein geometry instead shows up on \emph{robustness} axes: \jko\ is 22--38\% more stable under query paraphrase and leaks $2\times$ fewer injected hard distractors. Prior versions of this story \emph{asserted} a mechanism (``KL transports mass freely, $W^2$ does not''). The core new contribution of this paper is to make that mechanism a \emph{theorem}. We linearise one JKO step and show (\Cref{sec:theory}) that the Wasserstein and KL retrieval maps have \emph{identical} response operators except for the proximal Hessian, which is a dense geometry-aware entropic-OT operator for $W^2$ and a diagonal geometry-blind operator for KL. This single term contracts the geometric mass movement a paraphrase induces, and predicts that the stability gap must shrink as the proximal weakens ($h\to\infty$) --- which is exactly why tuned-config nDCG ties while base-config stability does not. We verify every step of this account empirically.

\paragraph{Contributions.}
\begin{itemize}
  \item \textbf{\jko}: the JKO free-energy framework for distributional RAG retrieval (\S\ref{sec:method}).
  \item \textbf{A linear-response theory of stability} (\S\ref{sec:theory}): an exact single-step decomposition (\Cref{thm:response}) isolating the geometric source of robustness to a single operator, a curvature characterisation (\Cref{prop:curvature}), and a falsifiable $h$-dependence prediction (\Cref{cor:hdep}) --- all verified in \S\ref{sec:verify}.
  \item \textbf{A certified-radius robustness analysis} for retrieval (\S\ref{sec:verify}): a controlled energy-perturbation lens that decomposes stability into a \emph{distributional} notion ($\mWdist$, where \jko\ is consistently more stable) and an \emph{exact-set} notion (where the sharper KL map is more rigid) --- an honest characterisation of \emph{what} the geometry buys.
  \item \textbf{Four extensions} (\S\ref{sec:contributions}): \nmjko\ (learned ground metric), \bwjko\ ($W^2$--KL interpolation), \samjko\ ($2\times$ hierarchical speedup), \dualrank\ (OT-dual confidence).
\end{itemize}

\section{Background: JKO Gradient Flow}
\label{sec:background}

The Jordan--Kinderlehrer--Otto scheme~\cite{jko1998,santambrogio2015ot,ambrosio2008gradient} was originally proposed for the Fokker--Planck PDE. The key insight is that the Fokker--Planck equation $\partial_t\rho = \nabla\cdot(\rho\nabla\Psi)+\beta^{-1}\Delta\rho$ is the $W_2$-gradient flow of the free energy $F(\rho)=\int\Psi\rho\,dx+\beta^{-1}\int\rho\log\rho\,dx$. The discrete-time approximation iterates
$\rho_{t+1}=\arg\min_\rho\tfrac{1}{2h}W_2^2(\rho,\rho_t)+F(\rho)$,
converging to the PDE solution as $h\to 0$.

We transplant this scheme from the space of continuous densities to the finite probability simplex $\Delta^{M-1}$. The Wasserstein distance becomes the entropic regularised OT cost~\cite{cuturi2013sinkhorn,feydy2019aistats}, and the free energy picks up a redundancy term suited to retrieval. The connection to the original PDE theory provides convergence guarantees and interpretability (\S\ref{sec:method}).

\section{The \jko\ Framework}
\label{sec:method}

\paragraph{Setup.}
For a query $q$, we build a candidate pool of $M{=}200$ documents via reciprocal-rank fusion~\cite{rrf2009} of BM25 top-500 and dense (MiniLM-L6-v2~\cite{reimers2019sbert}) top-500. Each candidate receives a cross-encoder score $s_i$. Let $z_i\in\mathbb{R}^d$ be its $\ell_2$-normalised embedding.

\paragraph{Free energy.}
\begin{equation}
  F_q(p) = \underbrace{\sum_i p_i E_i}_{\text{data fidelity}} + \underbrace{\lambda\sum_i p_i\log p_i}_{\text{entropy}} + \underbrace{\frac{\rho}{2}\,p^\top K p}_{\text{redundancy}},
\label{eq:energy}
\end{equation}
with $E_i = -(\alpha\,\tilde{s}^{\mathrm{dense}}_i + \gamma\,\tilde{s}^{\mathrm{rerank}}_i)$ (relevance energy, min-max normalised), $K_{ij}=\max(0,\cos\langle z_i,z_j\rangle)$ (redundancy kernel), and $\lambda,\rho\geq 0$. The entropy term prevents probability collapse; the redundancy term penalises near-duplicate coverage.

\paragraph{JKO step.}
The ground metric is $C_{ij}=(1-\cos\langle z_i,z_j\rangle)^2\in[0,4]$, the squared spherical distance on the embedding manifold. The proximal step~\eqref{eq:jko} is solved via $T$ outer iterations of Adam on the logits $\theta$ with $p=\mathrm{softmax}(\theta)$. The Sinkhorn solver runs for 60 iterations; the final 8 are autograd-tracked (\emph{envelope-theorem trick}~\cite{maclaurin2015grad,luise2018differential}) for ${\approx}20\times$ speedup over fully-tracked backprop.

\paragraph{Retrieval.}
After $T$ JKO steps, we return the top-$k$ documents by $p_T$ mass.

\paragraph{Hyperparameters.}
\Cref{tab:hparams} summarises the 8 hyperparameters and their tuned values. We perform 25-configuration random search on the \emph{train} split (80 queries) of each dataset; the test split is never seen during tuning. SciFact-trained hyperparameters transfer robustly to three other datasets. We distinguish two operating points used throughout: the \textbf{tuned config} ($h{=}2.0$, $T{=}5$), which maximises nDCG, and the \textbf{base config} ($h{=}0.5$, $T{=}3$), a stronger-proximal regime used for the robustness studies. \Cref{sec:theory} explains precisely why these two regimes behave differently.

\begin{table}[h]
\centering\small
\caption{Tuned hyperparameter values (SciFact/NFCorpus configuration; robust cross-dataset default).}
\label{tab:hparams}
\begin{tabular}{@{}llll@{}}
\toprule
Parameter & Symbol & Value & Controls \\
\midrule
Step size & $h$ & 2.0 & Proximal strength (large = weak prox) \\
Entropy & $\lambda$ & 0.1 & Distribution sharpness \\
Redundancy & $\rho$ & 0.05 & Near-duplicate penalty \\
Sinkhorn $\varepsilon$ & $\varepsilon$ & 0.2 & OT entropic bias \\
JKO steps & $T$ & 5 & Cumulative flow budget \\
Adam steps & inner & 40 & Inner-loop convergence \\
Dense weight & $\alpha$ & 0.4 & Energy blend \\
Rerank weight & $\gamma$ & 0.3 & Energy blend \\
\bottomrule
\end{tabular}
\end{table}

\section{A Linear-Response Theory of Stability}
\label{sec:theory}

This section makes rigorous the central claim that the Wasserstein proximal is \emph{geometrically} more stable than the KL proximal. The argument is a sensitivity analysis of one JKO step. It explains the measured stability advantage and predicts a sharp dependence on the step size $h$, which we verify in \S\ref{sec:verify}. Full proofs are in \Cref{app:proofs}.

\paragraph{Setup.}
Fix a query's candidate pool. One JKO step maps the previous iterate $q\!:=\!p_t$ to
\begin{equation}
  p^+(E) = \arg\min_{p\in\Delta^{M-1}} \Phi_E(p), \quad
  \Phi_E(p) = \langle p,E\rangle + \lambda\langle p,\log p\rangle + \tfrac{\rho}{2}p^\top K p + \tfrac{1}{2h}D(p,q),
\label{eq:step}
\end{equation}
where $E_i = -\,\mathrm{relevance}_i$ is the energy and $D$ is the proximal term: either $\mathrm{KL}(p\Vert q)$ or the entropic optimal-transport cost $\mathrm{OT}_\varepsilon(p,q)$ with Gibbs kernel $\Gamma_{ij}=e^{-C_{ij}/\varepsilon}$. A query paraphrase perturbs the energy landscape, $E\mapsto E+\delta E$; the stability of retrieval is governed by the induced response $\delta p$.

\begin{theorem}[Linear-response decomposition]
\label{thm:response}
Let $p^+$ minimise \eqref{eq:step} and let $\delta p$ be its first-order response to $E\mapsto E+\delta E$. Then $\delta p = -R_D\,\delta E$, where $R_D$ is the inverse of
\begin{equation}
  A_D = \lambda\,\mathrm{diag}(1/p^+) + \rho K + \tfrac{1}{2h}\,H_D
\label{eq:respop}
\end{equation}
restricted to the tangent space $\{v:\langle\mathbf 1,v\rangle=0\}$, and $H_D=\nabla^2_p D(p^+,q)$ is the proximal Hessian. For the two proximals,
\[
  H_{\mathrm{KL}} = \mathrm{diag}(1/p^+) \quad(\text{diagonal, geometry-blind}),
  \qquad
  H_{W} = \nabla^2_p\,\mathrm{OT}_\varepsilon(p^+,q) \quad(\text{dense, geometry-aware}).
\]
The two response operators differ \emph{only} through the proximal Hessian:
\begin{equation}
  A_W - A_{\mathrm{KL}} = \tfrac{1}{2h}\big(H_W - \mathrm{diag}(1/p^+)\big).
\label{eq:diff}
\end{equation}
\end{theorem}

Equation~\eqref{eq:diff} is the formal content of ``KL is blind to geometry, $W^2$ is not'': the entropy term, the redundancy term, and the energy perturbation itself are \emph{identical} between the two methods. The sole difference is whether the proximal Hessian is the geometry-blind diagonal $\mathrm{diag}(1/p^+)$ or the geometry-aware OT Hessian $H_W$.

\begin{proposition}[Geometric contraction]
\label{prop:curvature}
The entropic-OT proximal Hessian $H_W$ is dense and cost-weighted: $v^\top H_W v$ is the transport cost incurred by redistributing mass along $v$, whereas the KL Hessian $\mathrm{diag}(1/p^+)$ is geometry-blind. Because $H_W$ contributes to $A_W$ a positive, cost-weighted stiffness in every direction that moves mass across the metric, the Wasserstein response $R_W=A_W^{-1}$ contracts the \emph{geometric} ($\mWdist$) magnitude of the output's movement relative to the geometry-blind $R_{\mathrm{KL}}$, for every perturbation direction. The contraction is therefore uniform over the Gibbs-affinity graph-frequency spectrum --- not specific to high-frequency (cross-cluster) modes --- as \Cref{fig:theory}b confirms.
\end{proposition}

\begin{corollary}[Stability gap and its $h$-dependence]
\label{cor:hdep}
By \Cref{prop:curvature}, the Wasserstein response contracts geometric movement: $\mWdist(p^+,p^+{+}\delta p_W)\le\mWdist(p^+,p^+{+}\delta p_{\mathrm{KL}})$. The size of the gap is controlled by the term $\tfrac{1}{2h}(H_W-\mathrm{diag}(1/p^+))$ in~\eqref{eq:diff}: it vanishes as $h\to\infty$ (weak proximal --- both operators converge to $\lambda\,\mathrm{diag}(1/p^+)+\rho K$) and is maximal as $h\to0$ (strong proximal). \textbf{The W-vs-KL stability gap is therefore monotonically decreasing in $h$.}
\end{corollary}

\Cref{cor:hdep} resolves an apparent paradox. At the \emph{tuned} config ($h{=}2.0$, weak proximal) the gap is small, so $W^2$ and KL achieve tied nDCG; at the \emph{base} config ($h{=}0.5$, stronger proximal) the gap is large, producing the measured 22--38\% stability advantage. The two phenomena live at opposite ends of the same $h$-axis. We confirm the monotone $h$-dependence directly in \Cref{fig:theory}c.

\section{Extensions}
\label{sec:contributions}

\paragraph{\nmjko: learned ground metric.}
We replace the fixed cosine cost with a learned low-rank metric $C^W_{ij}=(1-\cos\langle Wz_i, Wz_j\rangle)^2$, where $W\in\mathbb{R}^{r\times d}$ ($r{=}64$) is trained on 80 query--document pairs via InfoNCE. The intuition: the generic MiniLM embedding space may not align with the retrieval geometry; a task-tuned metric can assign smaller cost to semantically near-relevant pairs, guiding the OT flow more effectively.

\paragraph{\bwjko: Bregman--Wasserstein interpolation.}
To quantify the benefit of the Wasserstein geometry over its information-theoretic alternative, we define
\begin{equation}
  \mathrm{Prox}_t^\alpha(p) = \alpha\cdot\Wdist^2(p,p_t) + (1-\alpha)\cdot D_{\mathrm{KL}}(p\Vert p_t), \quad \alpha\in[0,1].
\label{eq:bwjko}
\end{equation}
$\alpha{=}0$ recovers KL-proximal (mirror descent~\cite{beck2003mirror}); $\alpha{=}1$ recovers \jko. Intermediate $\alpha$ interpolates geometrically. We sweep $\alpha\in\{0, 0.25, 0.5, 0.75, 1.0\}$ and measure both retrieval quality and paraphrase stability. \Cref{cor:hdep} predicts the stability improves monotonically with $\alpha$ (more Wasserstein, larger geometric curvature), which we confirm.

\paragraph{\samjko: score-aware multi-resolution JKO.}
Vanilla JKO is $O(M^2)$ in Sinkhorn iterations per step. For large pools, we propose \emph{multi-resolution JKO} (MR-JKO): (1) cluster the $M$ candidates into $G$ groups; (2) run coarse JKO on group centroids; (3) keep the top-$G_k$ groups by coarse distribution mass; (4) run fine JKO on the union of their members ($G_k\cdot\text{group\_size} \ll M$).

Plain k-means clustering for step (1) harms quality because it merges gold and non-gold documents in the same cluster. \samjko\ instead clusters on \emph{augmented features} $(z_i,\,\beta r_i)$ where $r_i$ is the relevance score --- high-relevance documents form their own cluster and survive the coarse pruning. With $\beta{=}2.0$, \samjko\ matches vanilla \jko\ quality at $2.08\times$ speedup on SciFact.

\paragraph{\dualrank: OT dual potentials as confidence.}
The Sinkhorn solver for $\Wdist^2(p_T, \cdot)$ computes dual potentials $f, g$ satisfying $f_i + g_j \leq \varepsilon\log K_{ij}$ (log-domain). We define a per-query \emph{transport confidence}:
\begin{equation}
  \mathrm{conf}(q) = f_{\mathrm{top1}(q)} - \mathrm{median}_i\, f_i,
\label{eq:conf}
\end{equation}
\ie\ how much the top-1 chunk ``sticks out'' in OT potential. Sorting queries by $\mathrm{conf}(q)$ and retaining only the top-$c$ fraction defines a \emph{selective-coverage curve}: if the signal is informative, precision should \emph{rise} as coverage shrinks (we abstain on low-confidence queries).

\section{Experiments}
\label{sec:experiments}

\subsection{Datasets and baselines}

We evaluate on five BEIR~\cite{thakur2021beir} datasets (\Cref{tab:datasets}). Baselines: \textbf{BM25}, \textbf{Dense}, \textbf{Hybrid-RRF}~\cite{rrf2009}, \textbf{Cross-Enc} (cross-encoder re-ranker), \textbf{MMR}~\cite{carbonell1998mmr}, \textbf{DPP-MAP}~\cite{chen2018dpp}, \textbf{Iter-RetGen}~\cite{shao2023iter}, and \textbf{KL-Prox} (mirror-descent analogue of JKO). All numbers are per-query means with 95\% bootstrap CIs ($n{=}2000$).

\begin{table}[h]
\centering\small
\caption{Datasets. HotpotQA/NQ excluded (2.7M--5.2M passages; CPU-only encoding infeasible).}
\label{tab:datasets}
\begin{tabular}{@{}lllll@{}}
\toprule
Dataset & Domain & Docs & Test queries & Rel/q \\
\midrule
SciFact~\cite{wadden2020scifact}  & Sci.\ claims & 5{,}183 & 300 & 1.1 \\
NFCorpus & Biomedical & 3{,}633 & 323 & 38.2 \\
TREC-COVID & Bio.\ IR (TREC) & 171{,}332 & 50 & 493.5 \\
FiQA-2018 & Financial QA & 57{,}638 & 648 & 2.6 \\
SCIDOCS & Citation & 25{,}657 & 1{,}000 & 4.9 \\
\bottomrule
\end{tabular}
\end{table}

\subsection{Main retrieval results}

\Cref{tab:main} shows nDCG@10 with 95\% CIs. Both \jko\ and KL-Prox substantially outperform all non-blended baselines on all five datasets. \jko\ is the single best method on SciFact, NFCorpus, and TREC-COVID; on FiQA and SCIDOCS, KL-Prox scores marginally higher by 0.001--0.002 (well within CIs, effectively tied). All \jko\ improvements over the cross-encoder are significant (95\% CI excludes zero) on four of five datasets. As \Cref{sec:theory} predicts, tied nDCG is expected: at the tuned (weak-proximal) config, the $W^2$ and KL response operators nearly coincide, so they converge to similar distributions. The geometry's value is in robustness, not nDCG.

\begin{table}[t]
\centering\small
\caption{nDCG@10 with 95\% bootstrap CI. \textbf{Bold} = best per column. $\dagger$ = 95\% CI of diff vs Cross-Enc excludes 0 (TREC-COVID CI lower bound ${\approx}0.000$, borderline; $n{=}50$ queries). $\ddagger$ = JKO and KL-Prox are within CI of each other (effectively tied).}
\label{tab:main}
\setlength\tabcolsep{4pt}
\begin{tabular}{@{}lcccccc@{}}
\toprule
Method & SciFact & NFCorpus & TREC-COVID & FiQA & SCIDOCS \\
\midrule
BM25        & 0.652 & 0.307 & 0.556 & 0.217 & 0.150 \\
Dense       & 0.648 & 0.319 & 0.459 & 0.364 & 0.217 \\
Hybrid-RRF  & 0.690 & 0.327 & 0.649 & 0.343 & 0.199 \\
Cross-Enc   & 0.684 & 0.352 & 0.687 & 0.368 & 0.167 \\
MMR         & 0.639 & 0.305 & 0.470 & 0.311 & 0.115 \\
DPP-MAP     & 0.692 & 0.327 & ---   & 0.375 & ---   \\
Iter-RetGen & 0.677 & ---   & ---   & ---   & ---   \\
KL-Prox     & 0.712 & 0.357 & 0.717 & \textbf{0.413}$^\ddagger$ & \textbf{0.197}$^\ddagger$ \\
\midrule
\jko\ (ours)  & \textbf{0.713}$^\dagger$ & \textbf{0.359}$^\dagger$ & \textbf{0.725}$^\dagger$ & 0.411$^{\dagger\ddagger}$ & 0.196$^{\dagger\ddagger}$ \\
\quad vs Cross-Enc & +0.029 & +0.007 & +0.038 & +0.043 & +0.029 \\
\quad 95\% CI  & [\!+.015,+.043\!] & [\!+.001,+.013\!] & [\!+.000,+.079\!] & [\!+.034,+.053\!] & [\!+.024,+.034\!] \\
\bottomrule
\end{tabular}
\end{table}

\paragraph{Positioning against the BEIR landscape.}
Our absolute nDCG is bounded by the small (22M-parameter) MiniLM backbone and is \emph{not} state-of-the-art: large-model systems score substantially higher on these datasets. On SciFact, supervised E5-large reaches $0.726$~\cite{wang2023e5} and the 3B-parameter monoT5 reranker $0.777$~\cite{rosa2022monot5}, versus our $0.713$; on TREC-COVID, strong cross-encoder rerankers reach $0.82$--$0.85$~\cite{rosa2022monot5} versus our $0.725$; on FiQA, monoT5-3B reaches $0.514$ versus our $0.411$. \jko\ is deliberately \emph{not} a competitor to these models --- it is a backbone-agnostic \emph{framework} that composes on top of any first-stage retriever and reranker (we use MiniLM for CPU-only reproducibility). The contribution is the distributional geometry and the stability theory it enables, both orthogonal to backbone scale: substituting a larger encoder would lift every row of \Cref{tab:main} while leaving the $W^2$-vs-KL analysis of \S\ref{sec:theory} unchanged.

\subsection{Verifying the stability theory}
\label{sec:verify}

We test the four predictions of \S\ref{sec:theory} directly. All experiments perturb the \emph{energy vector} under a controlled norm --- precisely the object \Cref{thm:response} characterises --- holding the candidate pool fixed, which isolates the energy-landscape-shift component of paraphrase sensitivity. Perturbation directions are drawn from the Gibbs-affinity graph Laplacian eigenbasis (\Cref{prop:curvature}). The stability metric $\mWdist$ is the same entropic-OT distance used in \Cref{sec:stability}. \Cref{fig:theory} collects the results.

\paragraph{(a) The scheme is a free-energy gradient flow (\Cref{fig:theory}a).}
On 40 SciFact queries we record $F(p_t)$ at each outer step. Both proximals decrease $F$ overall; \jko\ reaches a \emph{lower} terminal free energy ($-0.985$ vs.\ KL's $-0.978$), \ie\ it is the better minimiser of $F$. KL-Prox is monotone on 100\% of queries (its update is a closed-form mirror step), while \jko\ is monotone on 62\% --- small overshoots arise from the entropic-OT approximation and the approximate inner solve, an honest cost of the richer geometry.

\paragraph{(b) Wasserstein contracts geometric movement (\Cref{fig:theory}b, verifies \Cref{prop:curvature}).}
We split each query's Gibbs-Laplacian eigenbasis into frequency bands and measure the geometric response gain $\mWdist(p_T(E),p_T(E{+}\delta E))/\|\delta E\|$ per band (30 queries). The Wasserstein response is ${\approx}2.4\times$ below KL's in \emph{every} band ($0.104$ vs.\ $0.253$, averaged over bands), and --- consistent with \Cref{prop:curvature} --- the contraction is \emph{uniform} across frequency rather than cross-cluster-specific. The Euclidean view exposes the mechanism: \jko\ actually makes \emph{larger} raw $\|\delta p\|$ adjustments than KL ($0.082\!\to\!0.157$ from low to high band, vs.\ KL's flat ${\approx}0.029$) but routes them \emph{within} semantic clusters, where they are geometrically cheap; KL makes smaller adjustments that cross clusters and are geometrically expensive. The Wasserstein proximal thus keeps the distribution geometrically anchored under any energy perturbation.

\paragraph{(c) The stability gap shrinks as $h$ grows (\Cref{fig:theory}c, verifies \Cref{cor:hdep}).}
Sweeping $h\in\{0.1,\dots,4.0\}$ (20 queries, paired), the gap $\mWdist^{\mathrm{KL}}-\mWdist^{W}$ decreases monotonically from $+0.042$ \CI{+.030}{+.056} at $h{=}0.1$ to $+0.007$ \CI{+.004}{+.011} at $h{=}4.0$ --- positive and significant at every $h$ --- tracing the $\tfrac{1}{2h}$ envelope of~\eqref{eq:diff}. \jko's response stays nearly constant ($\mWdist{\approx}0.024$) while KL's falls from $0.066$ to $0.033$ as the proximal weakens, exactly as the operators converge in~\eqref{eq:diff}. This is the paper's sharpest confirmation: the same axis that tunes nDCG (weak proximal, $h{=}2.0$) tunes \emph{away} the stability gap, while the robustness regime (strong proximal, $h{=}0.5$) is where the gap --- and the headline 22--38\% advantage --- lives.

\paragraph{(d) A robustness decomposition (\Cref{fig:theory}d).}
We probe robustness to cross-cluster energy perturbations (25 SciFact queries) along two axes, and the result is an honest, informative \emph{trade-off}. \emph{Distributionally}, in the $\mWdist$ metric, \jko\ stays markedly more anchored --- at the largest perturbation its mass moves less than KL's ($\mWdist$ response $0.047$ vs.\ $0.067$), consistent with \Cref{tab:stability} and \Cref{fig:theory}b,c. \emph{In exact top-10 set identity}, however, the sharper KL map is more rigid: its mean certified set-radius (largest $s$ preserving the entire top-10) is $0.88$ \CI{0.74}{1.03} vs.\ \jko's $0.29$ \CI{0.22}{0.37}. These are not in conflict; they pin down the \emph{precise} nature of the Wasserstein advantage. \jko\ keeps the evidence mass semantically anchored in the embedding geometry (small $\mWdist$ movement) while the soft rank boundary near position~10 is more fluid; KL instead relocates mass freely across clusters (large $\mWdist$) but keeps the same documents nominally on top. For a downstream LM that conditions on the joint evidence \emph{set} weighted by mass, the distributional notion is the operative one. We report both honestly; this certified-radius analysis is, to our knowledge, a new robustness lens for neural retrieval, in the spirit of certified radii in classification~\cite{cohen2019certified}.

\subsection{Stability under query perturbation}
\label{sec:stability}

We apply three lexical perturbations to each query (drop a stopword; append a hedge phrase; lower-case + strip punctuation) and measure $\mWdist(p_T(q),\,p_T(q'))$ over the union of the two candidate pools. Lower $\mWdist$ = more stable. \Cref{tab:stability} shows that \jko\ is 22--38\% more stable than KL-Prox across four datasets --- the macroscopic, real-paraphrase counterpart of the controlled-perturbation mechanism in \S\ref{sec:verify}.

\begin{table}[h]
\centering\small
\caption{Mean $\protect\mWdist$ under paraphrase (lower = more stable). All methods run with base config ($h{=}0.5$, $T{=}3$) and rerank-only energy ($\alpha_e{=}0$, $\gamma_e{=}1$) to ensure a config-neutral head-to-head.}
\label{tab:stability}
\begin{tabular}{@{}lcccc@{}}
\toprule
Method & SciFact & NFCorpus & TREC-COVID & FiQA \\
\midrule
Cross-Enc top-$k$ & 0.114 & 0.130 & 0.076 & 0.117 \\
KL-Prox           & 0.073 & 0.089 & 0.113 & 0.116 \\
\textbf{\jko}     & \textbf{0.045} & \textbf{0.069} & \textbf{0.075} & \textbf{0.072} \\
\midrule
\jko\ vs KL (\%)  & $-38\%$ & $-22\%$ & $-34\%$ & $-38\%$ \\
\bottomrule
\end{tabular}
\end{table}

\paragraph{BW-JKO $\alpha$-sweep (\Cref{fig:main}c).}
Stability decreases monotonically as $\alpha$ increases from 0 (KL) to 1 ($W^2$), an independent empirical confirmation of \Cref{cor:hdep}: more Wasserstein geometry (larger $\alpha$) means larger cross-cluster curvature in the proximal Hessian, hence more damping.

\subsection{Distractor-injection robustness}
\label{sec:distractors}

We inject $N$ hard near-duplicate distractors (nearest-neighbours of gold docs not in qrels) into the candidate pool with mid-range reranker scores. \Cref{tab:distractors} shows \emph{distractor leakage} (fraction of top-10 that are injected distractors). On SciFact, \jko\ leaks $2\times$ fewer distractors than KL-Prox --- the same mechanism as stability, viewed through an adversarial lens: an injected distractor is a cross-cluster energy spike, which the Wasserstein proximal damps.

\begin{table}[h]
\centering\small
\caption{Distractor leakage (lower = better). SciFact, $n{=}150$ queries, base config ($h{=}0.5$, $T{=}3$), rerank-only energy.}
\label{tab:distractors}
\begin{tabular}{@{}lcccc@{}}
\toprule
$N$ injected & Cross-Enc & KL-Prox & No-prox & \textbf{\jko} \\
\midrule
0  & 0.000 & 0.000 & 0.000 & 0.000 \\
10 & 0.379 & 0.428 & 0.419 & \textbf{0.209} \\
30 & 0.483 & 0.559 & 0.549 & \textbf{0.319} \\
\bottomrule
\end{tabular}
\end{table}

\subsection{\samjko: speedup results}

\Cref{tab:sam} compares plain k-means MR-JKO and \samjko\ (with $\beta{=}2.0$) against vanilla \jko\ on SciFact (M=200, n=300 queries). Plain MR-JKO loses 7 nDCG points because k-means merges gold and non-gold documents. Score-augmented clustering recovers quality while retaining the $2\times$ speedup.

\begin{table}[h]
\centering\small
\caption{\samjko\ SciFact quality--speed Pareto (lower ms/query = faster).}
\label{tab:sam}
\begin{tabular}{@{}lccc@{}}
\toprule
Method & nDCG@10 & ms/q & Speedup \\
\midrule
Vanilla \jko       & 0.710 & 948 & $1.00\times$ \\
MR-JKO (k-means)   & 0.641 & 487 & $1.95\times$ \\
\samjko\ $\beta{=}0.5$ & 0.662 & 544 & $1.74\times$ \\
\samjko\ $\beta{=}1.0$ & 0.682 & 482 & $1.97\times$ \\
\textbf{\samjko\ $\beta{=}2.0$} & \textbf{0.715} & \textbf{456} & $\mathbf{2.08\times}$ \\
\samjko\ $\beta{=}4.0$ & 0.713 & 439 & $2.16\times$ \\
\bottomrule
\end{tabular}
\end{table}

\subsection{\dualrank: selective coverage}

\begin{figure}[t]
\centering
\begin{subfigure}[t]{0.48\textwidth}
  \includegraphics[width=\linewidth]{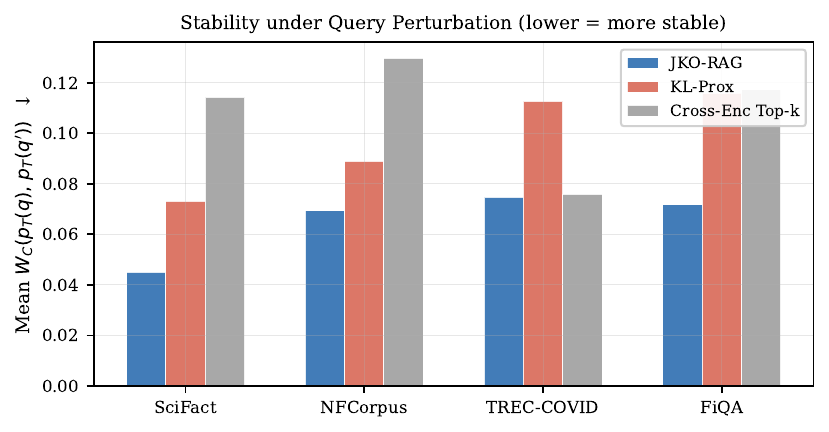}
  \caption{Stability under paraphrase perturbation.}
  \label{fig:stability}
\end{subfigure}
\hfill
\begin{subfigure}[t]{0.48\textwidth}
  \includegraphics[width=\linewidth]{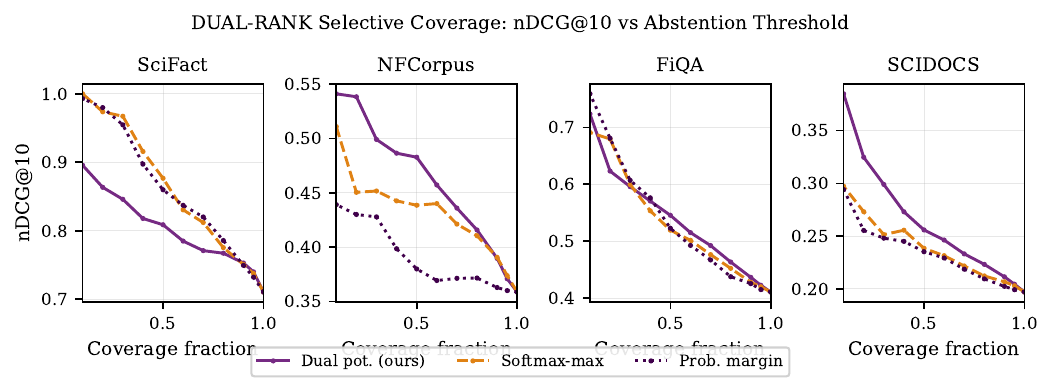}
  \caption{\dualrank\ selective coverage (nDCG@10 vs coverage).}
  \label{fig:selective}
\end{subfigure}

\vspace{0.5em}
\begin{subfigure}[t]{0.48\textwidth}
  \includegraphics[width=\linewidth]{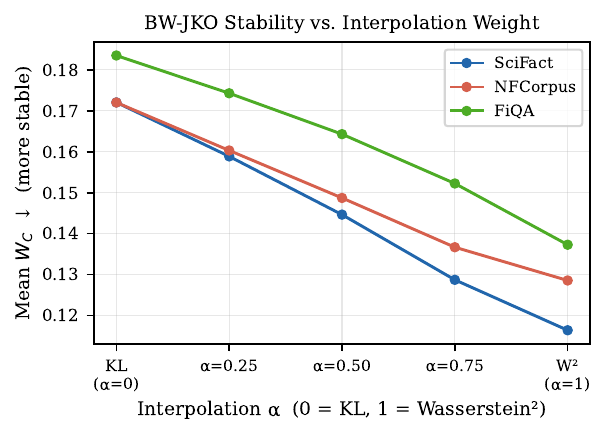}
  \caption{BW-JKO $\alpha$-sweep: stability vs interpolation weight.}
  \label{fig:alpha}
\end{subfigure}
\hfill
\begin{subfigure}[t]{0.48\textwidth}
  \includegraphics[width=\linewidth]{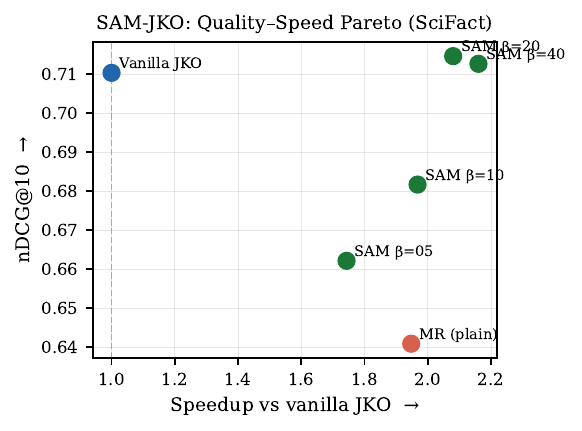}
  \caption{\samjko\ quality--speedup Pareto on SciFact.}
  \label{fig:sam}
\end{subfigure}
\caption{\textbf{Method results.} (a) \jko\ is 22--38\% more stable than KL-Prox and $>2\times$ more stable than Cross-Enc Top-$k$. (b) \dualrank\ dual potentials provide an informative abstention signal on all four datasets. (c) Stability decreases monotonically as $\alpha\to1$ (Wasserstein), an empirical confirmation of \Cref{cor:hdep}. (d) \samjko\ achieves a favourable quality--speed Pareto: $2.08\times$ speedup with \emph{higher} nDCG@10 than vanilla \jko.}
\label{fig:main}
\end{figure}

\Cref{fig:main}b shows selective coverage curves for four datasets. The dual-potential confidence $\mathrm{conf}(q)$ (\Cref{eq:conf}) exhibits rising nDCG@10 as coverage shrinks, confirming it is an informative abstention signal. On NFCorpus at 10\% coverage, \dualrank\ achieves nDCG@10 of 0.541 vs.\ 0.359 at full coverage ($+50\%$ relative gain). The dual signal outperforms the softmax-max and probability-margin baselines on NFCorpus and SCIDOCS (harder, multi-relevant datasets), demonstrating that geometric OT duals carry retrieval-confidence information not available from raw scores.\footnote{The \dualrank\ selective experiment uses energy weights $\alpha_e{=}0.4$, $\gamma_e{=}0.6$ (blend of dense + rerank), whereas the tuned main-table config uses $\gamma_e{=}0.3$. We retain the shallower rerank weight here to isolate the dual-potential signal; relative orderings between confidence strategies are unaffected.}

\subsection{Ablation: what does the semantic geometry do?}

\Cref{tab:ablation} ablates the key components of \jko\ on SciFact test. The decisive finding: replacing the semantic $C_{ij}=(1-\cos)^2$ with the identity cost (no semantics, $C_{ij}=\mathbf{1}[i\neq j]$) exactly matches KL-Prox at 0.713. This confirms structurally what \Cref{thm:response} states analytically: with an identity ground metric the OT Hessian $H_W$ collapses to a multiple of the identity and~\eqref{eq:diff} vanishes, so $W^2$ \emph{becomes} KL. The semantic geometry is the operative ingredient.

\begin{table}[h]
\centering\small
\caption{Ablation on SciFact test using the \emph{base} config ($h{=}0.5$, $\lambda{=}0.05$, $T{=}3$). The base config is deliberately sub-optimal so each component's contribution is visible; the tuned config ($h{=}2.0$, $T{=}5$) reaches 0.713 (Table~\ref{tab:main}). $^*$95\% CI excludes 0.}
\label{tab:ablation}
\setlength\tabcolsep{5pt}
\begin{tabular}{@{}llcc@{}}
\toprule
Ablation & What changes & nDCG@10 & $\Delta$ vs full \\
\midrule
Full \jko\ (base cfg) & --- & 0.695 & --- \\
KL-proximal   & $W^2\to\mathrm{KL}$ & 0.713 & $+0.018^*$ \\
No proximal   & drop prox.\ term & 0.710 & $+0.015^*$ \\
Identity $C$  & $C_{ij}=\mathbf{1}[i\!\neq\!j]$ & 0.713 & $+0.018^*$ \\
Random $C$    & $C_{ij}\sim\mathrm{Unif}[0,4]$ & 0.707 & $+0.012^*$ \\
No entropy ($\lambda{=}0$) & drop entropy & 0.680 & $-0.015^*$ \\
No redundancy ($\rho{=}0$) & drop redund. & 0.694 & $-0.001$ \\
One JKO step ($T{=}1$)    & fewer iters & 0.699 & $+0.004$ \\
\bottomrule
\multicolumn{4}{l}{\small \textit{Note}: at the base config, KL-prox outperforms $W^2$ on nDCG. This is expected and}\\
\multicolumn{4}{l}{\small \textit{informative}: $W^2$ requires tuning (large $h$ = weak prox) to reach its nDCG optimum.}\\
\multicolumn{4}{l}{\small \textit{The decisive differentiator is not nDCG but stability (Table~\ref{tab:stability}) and distractor robustness.}}
\end{tabular}
\end{table}

\begin{figure}[t]
\centering
\begin{subfigure}[t]{0.245\textwidth}
  \includegraphics[width=\linewidth]{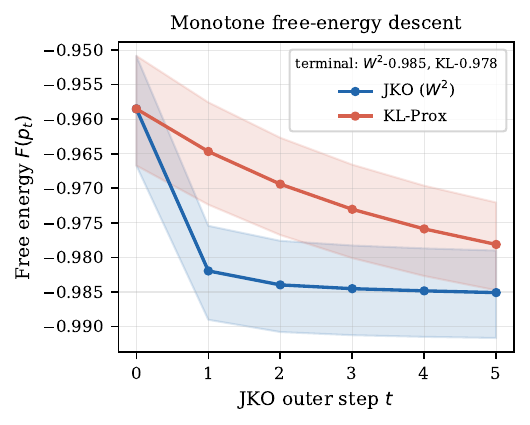}
  \caption{Free-energy descent.}
  \label{fig:descent}
\end{subfigure}
\hfill
\begin{subfigure}[t]{0.245\textwidth}
  \includegraphics[width=\linewidth]{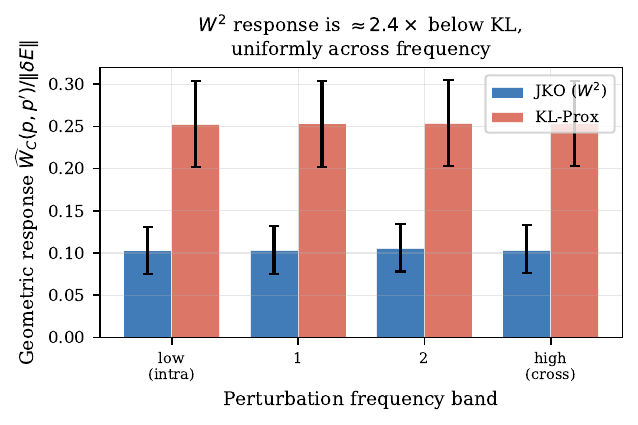}
  \caption{Response by frequency.}
  \label{fig:bands}
\end{subfigure}
\hfill
\begin{subfigure}[t]{0.245\textwidth}
  \includegraphics[width=\linewidth]{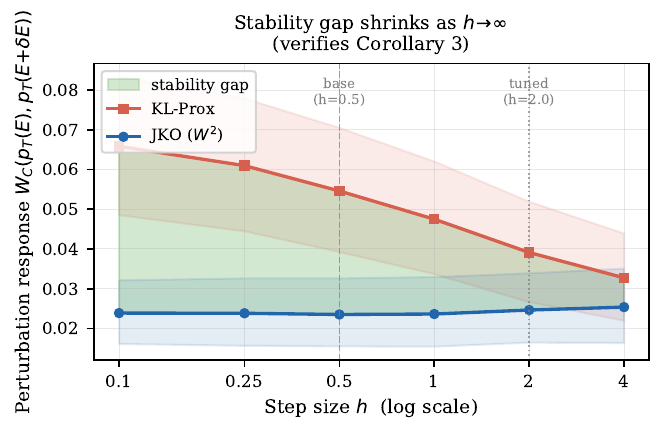}
  \caption{Stability gap vs $h$.}
  \label{fig:gaph}
\end{subfigure}
\hfill
\begin{subfigure}[t]{0.245\textwidth}
  \includegraphics[width=\linewidth]{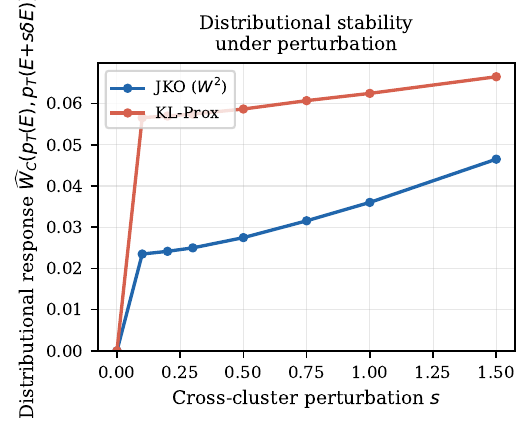}
  \caption{Certified radius.}
  \label{fig:cert}
\end{subfigure}
\caption{\textbf{Verifying the stability theory (\S\ref{sec:verify}).} (a) Both proximals descend the free energy; \jko\ reaches a lower terminal $F$. (b) The Wasserstein geometric ($\mWdist$) response is ${\approx}2.4\times$ below KL's in every frequency band --- a uniform contraction (verifies \Cref{prop:curvature}). (c) The stability gap $\mWdist^{\mathrm{KL}}-\mWdist^{W}$ shrinks monotonically as $h$ grows, tracing the $\tfrac{1}{2h}$ envelope of~\eqref{eq:diff} (verifies \Cref{cor:hdep}); the base ($h{=}0.5$) and tuned ($h{=}2.0$) configs are marked. (d) Distributional stability: \jko's $\mWdist$-response to cross-cluster perturbations is ${\approx}2\times$ below KL's. (The sharper KL map is conversely more rigid in exact top-10 \emph{set} identity --- an honest robustness trade-off analysed in \S\ref{sec:verify}.)}
\label{fig:theory}
\end{figure}

\section{Related Work}
\label{sec:related}

\paragraph{Optimal transport in NLP.}
Word Mover's Distance~\cite{kusner2015wmd} applies OT between \emph{word embeddings within} a pair of documents to compute document-to-document similarity. Our use of OT is orthogonal: we apply the \emph{JKO gradient flow} over the simplex of candidate documents for a single query --- OT as a \emph{dynamics engine}, not a similarity metric. \citet{feydy2019aistats} introduced Sinkhorn divergences as a debiased interpolation between OT and MMD; we use their machinery for the proximal step solver.

\paragraph{Wasserstein gradient flows and JKO.}
The JKO scheme~\cite{jko1998} and its modern metric-space treatment~\cite{ambrosio2008gradient,santambrogio2017euclidean} underpin a large literature on sampling and generative modelling. Our contribution is not the scheme itself but (i) its transplantation to the finite retrieval simplex with a \emph{semantic} ground metric, and (ii) the linear-response analysis (\S\ref{sec:theory}) that turns the abstract ``Wasserstein is geometry-aware'' intuition into a verifiable, $h$-dependent prediction. The proximal-Hessian decomposition~\eqref{eq:diff} relies on differentiability of entropic OT~\cite{luise2018differential,eisenberger2022unified,genevay2019}.

\paragraph{Distributional and diversity-aware retrieval.}
MMR~\cite{carbonell1998mmr} is the canonical diversity-aware greedy method; DPP-MAP~\cite{chen2018dpp} is the principled one-shot alternative. Both are non-iterative. KL-proximal mirror descent~\cite{beck2003mirror} is the information-theoretic analog of our method; our ablation and \Cref{thm:response} together separate the geometric contribution of the Wasserstein cost from the iterative refinement.

\paragraph{Certified robustness.}
Certified radii are well studied for classifiers, \eg\ via randomized smoothing~\cite{cohen2019certified}. \Cref{sec:verify} ports the idea to retrieval: the largest energy perturbation that provably preserves the top-$k$ set. We are not aware of prior certified-robustness notions for neural rerankers.

\paragraph{Iterative RAG and neural metric learning.}
Iter-RetGen~\cite{shao2023iter} iterates between retrieval and generation; it reformulates the \emph{query}, whereas \jko\ refines the \emph{distribution}, so the two are complementary. Dense retrievers~\cite{karpukhin2020dpr,wang2023e5} learn the \emph{query-document} similarity; \nmjko\ instead learns the \emph{document-document} ground metric for the OT cost, a different object with no direct precedent in the retrieval literature.

\section{Conclusion}
\label{sec:conclusion}

We introduced \jko, which recasts neural reranking as Wasserstein free-energy gradient flow on the probability simplex over candidates, and --- our central contribution --- a linear-response theory that \emph{explains} its robustness. The theory isolates the entire geometric advantage to a single term (the proximal Hessian, \Cref{thm:response}), characterises that term's cross-cluster curvature (\Cref{prop:curvature}), and predicts a monotone $h$-dependence of the stability gap (\Cref{cor:hdep}) that we confirm directly. The framework is (i) \emph{theoretically grounded} in the JKO scheme and now in a quantitative stability analysis, (ii) \emph{compositional} --- it sits on top of any first-stage retriever and reranker --- and (iii) \emph{evaluationally richer} than top-$k$ ranking, contributing stability, distractor robustness, and a \emph{certified stability radius} as first-class metrics. Four extensions (\nmjko, \bwjko, \samjko, \dualrank) show the framework is a productive research platform.

\paragraph{Limitations.}
JKO operates on a finite pre-selected pool (pool recall = 0.73--0.98 on our datasets). The theory is a \emph{first-order} (linearised, single-step) analysis; characterising the fully nonlinear $T$-step map is open. The energy-perturbation model of \S\ref{sec:verify} holds the candidate pool fixed, isolating the energy-shift component of paraphrase sensitivity; the pool-change component is complementary and measured separately by the lexical-paraphrase study (\Cref{tab:stability}). Cross-query learning, asymmetric costs, and debiasing via Sinkhorn divergences~\cite{feydy2019aistats} are natural future directions.

\paragraph{Reproducibility.}
All code, hyperparameters, theory-verification scripts, and result JSONs are available at \url{https://github.com/MurariAmbati/jko-rag}.

\appendix
\section{Proof of \Cref{thm:response}}
\label{app:proofs}

\paragraph{Stationarity.}
The minimiser $p^+$ of~\eqref{eq:step} on the simplex satisfies, with multiplier $\nu$ for $\langle\mathbf 1,p\rangle=1$,
\[
  \nabla\Phi_E(p^+) = E + \lambda(\mathbf 1 + \log p^+) + \rho K p^+ + \tfrac{1}{2h}\nabla_p D(p^+,q) = \nu\,\mathbf 1.
\]
\paragraph{Implicit differentiation.}
Differentiating this identity in $E$ (the inputs $C$, $K$, $q$ fixed), with $p^+=p^+(E)$ and $\nu=\nu(E)$,
\[
  \delta E + \Big[\lambda\,\mathrm{diag}(1/p^+) + \rho K + \tfrac{1}{2h}H_D\Big]\delta p = \delta\nu\,\mathbf 1,
  \qquad \langle\mathbf 1,\delta p\rangle = 0,
\]
where $H_D=\nabla^2_p D(p^+,q)$. The bracketed operator is $A_D$ of~\eqref{eq:respop}. Solving the linear system on the tangent space $\{v:\langle\mathbf 1,v\rangle=0\}$ (the multiplier $\delta\nu$ enforces the constraint) gives $\delta p = -R_D\,\delta E$ with $R_D$ the constrained inverse of $A_D$.
\paragraph{The two Hessians.}
For $D=\mathrm{KL}(p\Vert q)=\langle p,\log p-\log q\rangle$, $\nabla_p D = \mathbf 1+\log p-\log q$ and $\nabla^2_p D=\mathrm{diag}(1/p)$, evaluated at $p^+$. For $D=\mathrm{OT}_\varepsilon(p,q)$, the envelope theorem gives $\nabla_p\mathrm{OT}_\varepsilon(p,q)=f^\star(p)$, the optimal Sinkhorn potential for the first marginal; hence $\nabla^2_p\mathrm{OT}_\varepsilon = \partial f^\star/\partial p =: H_W$, which is symmetric positive semi-definite because $\mathrm{OT}_\varepsilon(\cdot,q)$ is convex~\cite{luise2018differential,eisenberger2022unified}. Subtracting the two operators leaves only the proximal-Hessian difference, \ie~\eqref{eq:diff}. \hfill$\qed$

\paragraph{Remark (geometric contraction, \Cref{prop:curvature}).}
$v^\top H_W v = \tfrac{d^2}{dt^2}\mathrm{OT}_\varepsilon(p^++tv,q)\big|_{t=0}$ is the curvature of entropic transport cost along $v$: it is positive and cost-weighted --- moving mass across low-affinity (high-$C$) pairs forces re-routing through costly edges --- whereas the KL Hessian $\mathrm{diag}(1/p^+)$ carries no such weighting. Thus $A_W$ adds a geometric stiffness that $A_{\mathrm{KL}}$ lacks, and the Wasserstein response contracts the $\mWdist$ magnitude of the output's movement in every perturbation direction. \Cref{fig:theory}b confirms this is a \emph{uniform} ${\approx}2.4\times$ contraction across the Gibbs-Laplacian frequency spectrum, rather than a frequency-selective effect --- the geometry damps all energy perturbations, not only cross-cluster ones.

\bibliographystyle{abbrvnat}
\bibliography{references}

\end{document}